# Broadening of hot-spot response spectrum of superconducting NbN nanowire single-photon detector with reduced nitrogen content

D. Henrich,[1a)] S. Dörner,[1] M. Hofherr,[1] K. Il'in,[1] A. Semenov,[2] E. Heintze,[3] M. Scheffler,[3] M. Dressel,[3] and M. Siegel[1]

[1]*Institut für Mikro- und Nanoelektronische Systeme, KIT, Hertzstr. 16, 76187 Karlsruhe, Germany*
[2]*DLR Institut für Planetenforschung, Rutherfordstr. 2, 12489 Berlin, Germany*
[3]*1. Physikalisches Institut, Universität Stuttgart, Pfaffenwaldring 57, 70550 Stuttgart, Germany*

The spectral detection efficiency and the dark count rate of superconducting nanowire single-photon detectors (SNSPD) has been studied systematically on detectors made from thin NbN films with different chemical compositions. Reduction of the nitrogen content in the 4 nm thick NbN films results in a more than two orders of magnitude decrease of the dark count rates and in a red shift of the cut-off wavelength of the hot-spot SNSPD response. The observed phenomena are explained by an improvement of uniformity of NbN films that has been confirmed by a decrease of resistivity and an increase of the ratio of the measured critical current to the depairing current. The latter factor is considered as the most crucial for both the cut-off wavelength and the dark count rates of SNSPD. Based on our results we propose a set of criteria for material properties to optimize SNSPD in the infrared spectral region.

**I. INTRODUCTION**

Superconducting nanowire single-photon detectors (SNSPD) have been intensively studied over the last decade[1]. The interest from numerous applications, where fast infrared detectors with ultimate sensitivity are required, is based on the promising results already demonstrated for SNSPD in the optical range[2]. Therefore efforts of research groups developing SNSPD are focused on the extension of the hot-spot response spectrum into the infrared range. It has been shown[3] that the minimal photon energy, which is required to generate a hot-spot of sufficient size, depends on the nanowire cross-section (i.e. thickness $d$ of the film used for the device fabrication and width $w$ of nanowire) and on material parameters like the superconducting energy gap $\Delta$, quasi-particle diffusion coefficient $D$, density of electron states $N_0$, time constant of quasi-particle multiplication $\tau$, the quasi-particle multiplication efficiency $\varsigma$, and the ratio of the bias current $I_B$ to the departing critical current $I_C^d$:

$$\lambda_C^{-1} = \frac{3\sqrt{\pi}}{4hc\varsigma} \Delta^2 w d N_0 \sqrt{D\tau} \left(1 - \frac{I_B}{I_C^d}\right) \qquad (1)$$

Obviously, according to (1) the cut-off wavelength $\lambda_C$ can be shifted towards lower energy infrared photons by reduction of the nanowire cross-section. It has been proven experimentally for SNSPD made

---
[a)] corresponding author e-Mail address: dagmar.rall@kit.edu

from NbN thin films on sapphire, that a decrease of the film thickness results not only in a red shift of $\lambda_c$, but also in an increase of the intrinsic detection efficiency of the nanowire[4]. Similarly, the reduction in NbN nanowire width below 50 nm has been demonstrated to broaden the spectral bandwidth into the infrared range[5]. However, this approach is limited first of all by the resolution of the employed lithography as well as the thin film deposition technique. Moreover, by reduction of the nanowire cross-section, the maximal bias current $I_B$ (limited by the critical current $I_C = j_C w d$) is decreased, which leads to smaller pulse amplitudes and consequently more demanding readout of low voltage signals. This problem can be partly solved by connecting nanowires of the detectors in parallel, where the switching in one strip due to a count event will induce an avalanche-like switching of the remaining strips[6]. However, this approach is limited itself by the afterpulsing that has been observed in 30 nm wide nanowire avalanche photodetectors[7].

The SNSPD spectrum also depends strongly on the ratio of the applied bias current to the pair-breaking current. While the latter is determined by the material, the maximum experimentally achievable critical current ($I_C^m$) in an SNSPD usually stays well below the theoretical limit of the pair-breaking current ($I_C^d$)[4]. One reason is the suppression of the critical current by the sharp turns of the nanowire of the detector usually made in form of a meander line[8]. Due to current piling at the inner radius of a bend nanowire, the critical current density is already locally reached at lower bias currents and causes a premature switch into the normal-conducting state. This effect scales with the ratio of width to coherence length. However, it has been shown experimentally that suppression of the measured critical current with respect to $I_C^d$ in both NbN[9] and NbTiN[10] thin film structures due to change of their geometry is much smaller than predicted by theory and probably caused by non-uniformity of these materials.

A third approach to increase $\lambda_c$ is to choose a superconducting material with smaller energy gap $\Delta$. The most widely used materials NbN, which can be deposited into very thin films below 4 nm[11], and NbTiN, which can be grown on silicon, enabling easy circuit integration and showing fewer dark counts[12], have typical critical temperature values of ~15 K. The dependence of the minimum required photon energy on the square of $\Delta$ should increase $\lambda_c$ values more than a factor of two with decrease in the superconducting energy by only 30% from 15 K to about 10 K. Several experimental investigations of different superconductors with lower $T_C$ values than the standard materials have been performed. Naturally, the first material of choice is pure Nb with bulk $T_C$ = 9.25 K, which has the additional benefit that nanowire detectors made from Nb have shown improved dead times compared to NbN[13]. However, they also demonstrate much lower detection efficiency and significantly shorter cut-off wavelength[14], in contrast to the expected broadening of the spectrum. It has been assumed that one reason for the obtained results is that the out-diffusion of quasi-particles from the hot-spot is faster in Nb nanowires than in NbN nanowires. Therefore, in Nb the collapse of the hot spot occurs on a shorter time scale than the redistribution of the applied current and the formation of a normal belt across the nanowire. Since the electron diffusion coefficient and the electron density of states are not independent parameters, it is convenient to compare square resistances $R_S = \rho/d$ ($\rho$ is the specific resistivity) of nanowires from different materials. In the framework of the free electron model, $R_S = N_0 d D e^2$. Hence the reciprocal cut-off wavelength appears to be proportional to $\varsigma^{-1}\Delta^2 w R_S (\tau/D)^{1/2}$. Consequently tantalum nitride, with $T_C$ slightly higher than in Nb but with the electron diffusion coefficient and the square resistance similar to that in NbN, has been suggested as an alternative material for SNSPD development[15]. Indeed, an increase of the cut-off wavelength by a factor ≈ 1.3 has been experimentally found for TaN SNSPD[16] with the same cross-section area like in NbN detectors in a good agreement with the model predictions (1). Conversely, for the detectors made from magnesium diboride, with relatively high $T_C$ values of 18 K, it was shown that infrared photons λ ≥ 1560 nm can no longer be detected[17].

Recently two types of Si compounds with the critical temperature values about or even below 2 K have been experimentally considered for development of SNSPD. Amorphous NbSi demonstrated higher DE above 1550 nm wavelengths relative to a NbTiN SNSPD[18]. DE values of 19%-40% in the wavelength region from 1280 nm to 1650 nm have been shown for devices made from amorphous tungsten silicide[19]. Although both materials can be deposited without heating of the substrate, the low $T_C$ confines the operation temperature range of the detectors made from NbSi and WSi into the sub-Kelvin range, limiting the area of their possible applications.

The superconducting energy gap can be varied not only by using different superconducting materials, but also by utilizing the proximity effect with normal metals, which is widely used for the adjustment of $T_C$ of ultra-sensitive transition edge sensors (TES) operating at millikelvin temperatures. Also, reducing the film thickness results in a decreased energy gap and correspondent $T_C$ values of superconducting films (as has been shown experimentally e.g. for NbN[20], TaN[15] and Nb[21]) when the thickness becomes comparable to the coherence length of the correspondent superconductor[22]. The unique feature of two components superconductors like nitrides and silicides of transition metals is that their parameters can be changed in a wide range from insulator to superconductor[23,24] by variation of the composition. Since the change in stoichiometry results not only in change of the superconducting energy gap value but also influence on almost all other material parameters it is not easy to predict in advance what composition will be optimal for the development of SNSPD with high efficiency to infrared photons.

In this paper, we present results of a systematic study of detection efficiency spectra and dark count rates of SNSPD made from 4 nm thick NbN films with varying chemical composition. We continuously change the material parameters of the NbN thin films by variation of deposition conditions (Section 1). SNSPD structures with identical geometry have been patterned from these films (Section 2) and their detector performance has been measured in a wide spectral range (Section 3). The shift of the cut-off wavelength towards longer wavelengths has been observed for the detectors made from NbN films with decreased nitrogen content. Material and geometrical factors responsible for the broadening of SNSPD spectra in these films are discussed in the last section of this paper.

**II. TECHNOLOGY OF NbN THIN FILMS AND THEIR CHARACTERIZATION**

**A. Reactive magnetron sputtering**

The NbN films were deposited on one-side polished R-plane $Al_2O_3$ substrates which were placed without any thermal glue on a heater kept at 750°C during deposition. The deposition was made by reactive magnetron sputtering of pure Nb target in an Argon and nitrogen atmosphere. Prior to deposition, the surface of the Nb target (99.95%) was pre-cleaned by sputtering of the target material in a pure Ar atmosphere at the pressure $p_{Ar} \approx 3\times10^{-3}$ mbar. The correspondent current-voltage characteristic (IV-curve) of the magnetron discharge taken in current bias mode is shown by open symbols in Fig. 1.

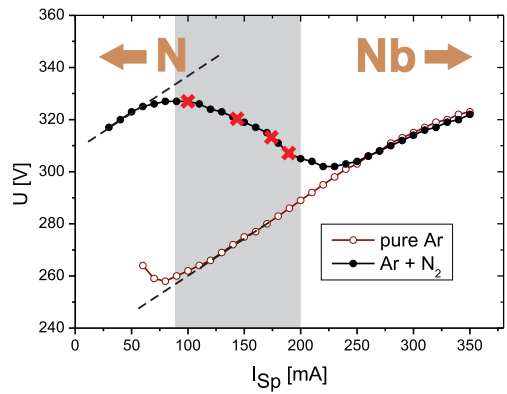

FIG. 1. (Color online) Discharge voltage characteristic of the plasma in the magnetron sputter chamber. The pure Argon plasma (empty circles) is stable for currents above 80 mA. When nitrogen is added to the atmosphere (filled circles), the dependence becomes non-monotonic with a similar but off-set dependence for low currents, illustrated by the two parallel dashed lines. In the transition region, superconducting NbN is deposited (grey area). The higher the sputter current $I_{SP}$, the higher the Nb content in the sputtered film. Red crosses mark the deposition conditions of the four selected films.

At currents above 80 mA the discharge voltage continuously rises with increasing discharge current. Reducing the current below 80 mA, the voltage increases drastically due to a reduced secondary electron emission rate at these conditions. After the pre-cleaning the reactive gas ($N_2$) with a partial pressure $P_{N2} \approx 2\times10^{-4}$ mbar was introduced into the vacuum chamber. This results in a significant change in the current-voltage characteristic of the discharge (Fig. 1, solid symbols) and it becomes non-monotonic, which is typical for reactive magnetron sputtering processes[25]. At currents between 20 to about 90 mA the discharge voltage increases with increase in current so that IV-curve of the discharge in Ar+$N_2$ atmosphere goes almost parallel to that in pure Ar (dashed lines in FIG. 1), but at about 70 V higher discharge voltage values. The difference in voltage is caused by a lower potential for the secondary ion-electron emission at the pure Nb surface of the target in comparison to the surface completely or partly covered by niobium nitride. After reaching a maximum at about 90 mA the voltage starts to decrease with increase in current up to about 225 mA. At this current value the discharge voltage becomes almost equal to the correspondent voltage of the discharge in the pure Ar atmosphere. At higher currents both curves (in Ar and in Ar+$N_2$ atmosphere) coincide with each other. The dependence of the discharge voltage on current in Ar+$N_2$ atmosphere can be qualitatively described with a competition of two simultaneous processes taking place on the surface of the Nb target. The first process is a chemical reaction between Nb and the radicalized nitrogen which leads to formation of niobium nitride layer on the target surface. The rate of this process is determined (at first approximation) by the partial pressure of $N_2$ and independent of discharge or in another word sputter current $I_{Sp}$. The second process is the sputtering of material from the target surface by accelerated $Ar^+$ ions. The sputter rate increases with increase of $I_{Sp}$. At low currents ($I_{Sp}$< 90 mA) the sputter rate is lower than the rate of the reaction process between Nb and nitrogen, thus the target surface is completely covered with NbN. At currents above 250 mA the situation is opposite and sputtering of the target surface is faster than formation of NbN on the surface leading to an almost nitride free surface of the target during deposition. The most interesting region for deposition of NbN films with optimal properties is the currents at which the discharge process is characterized by negative resistance (grey zone in Fig. 1). In this crossover region from nitrogen (low currents) to Nb (high currents) phase of discharge, the rates of both competing processes become equal to each other so that freshly formed NbN is immediately sputtered from the target surface. The resulting discharge voltage at this part of the IV-curve is determined by the fraction of the target surface covered with nitride, which decreases with increasing $I_{SP}$. NbN films deposited in the nitrogen phase are supposed to be with excess amount of nitrogen and the films deposited closer to the Nb phase are with reduced nitrogen content. The variation of stoichiometric composition and lattice formation of NbN films by change in the deposition parameters of reactive magnetron sputtering was confirmed by Bacon *et al.*[26] To ensure reproducibility and compatibility of the deposited films, the discharge of the mixture of Ar and $N_2$ gases was stabilized for 10-12 min at $I_{SP}$ = 100 mA. After that the current was slowly increased to the desired sputter value at which the film was deposited onto the substrate after additional 2-3 min stabilization of the discharge voltage. The deposition rate of NbN in the crossover region varies between 0.06 and 0.17 nm/s for 100 and 190 mA sputter current respectively.

**B. Critical temperature, $B_{C2}$ and *RRR* vs $I_{Sp}$**

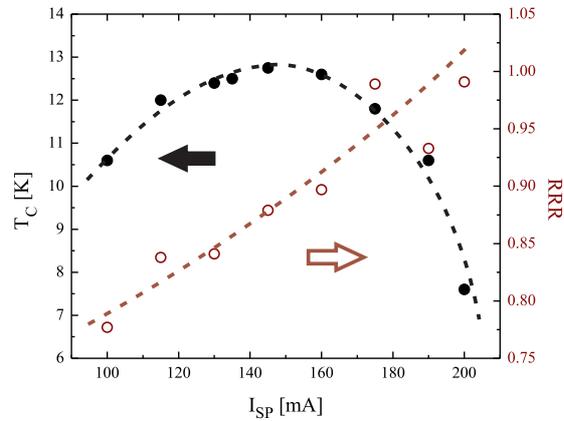

FIG. 2. (Color online) Parameters of 4 nm thick NbN films deposited at various sputter currents $I_{SP}$ covering the gray region of figure 1. The critical temperature $T_C$ (filled circles, left axis) shows an optimum at 145 mA. The residual resistivity ratio RRR (empty circles, right axis) increases continually with increasing sputter current. Dashed lines are to guide the eyes.

In Fig. 2 the solid symbols show the dependence of the critical temperature $T_C$ (defined as the temperature for which the resistance dropped below 1% of the normal state resistance $R_N$ within measurement accuracy) of 4.3±0.4 nm thick NbN films on sputter current. The maximum $T_C$ value has been reached for films deposited at 145 mA. Both increase and decrease of the sputter current from this value results in a reduced critical temperature and this reduction is stronger towards the Nb discharge phase. The sheet resistance above the superconducting transition $R_S$(20 K) decreases from 510 Ω for films deposited at the lowest $I_{Sp}$ to 254 Ω for those deposited at the highest $I_{Sp}$ (see Table I), confirming the assumption that with increasing $I_{Sp}$, the films have a higher Nb content and become more 'metal-like'. This is further confirmed by the increase of the residual resistivity ratio $RRR = R_S(300\,K)/R_S(20\,K)$ for films deposited at higher currents (see Fig. 2, open symbols). The dependence of the critical current density at 4.2 K on deposition current is similar to that of the critical temperature but with the maximum reached at higher $I_{Sp}$ = 175 mA. For further analysis we have selected four films from the crossover region deposited at 145 mA (maximum $T_C$), 175 mA (maximum $j_C$(4.2 K)), and two extremes with high nitrogen (100 mA) and high niobium content (190 mA). The last two have almost equal $T_C$ values.

TABLE I. Parameters of NbN thin films deposited at the sputter current $I_{SP}$: Thickness $d$, sheet resistance $R_S$ at 20 K, critical temperature $T_C$, critical current density $j_C$ at 4.2 K, electron diffusion coefficient $D$, coherence length $\xi$, energy gap $2\Delta$ at 4.2 K and calculated cut-off prefactor $\lambda_C(0)$.

| $I_{SP}$ [mA] | $d$ [nm] | $R_S$ [Ω] | $T_C$ [K] | $j_C$ [MA/cm$^2$] | $D$ [cm$^2$/s] | $\xi(0)$ [nm] | $2\Delta$ [meV] | $\lambda_C(0)$ [nm] |
|---|---|---|---|---|---|---|---|---|
| 100 | 4.7 | 510 | 10.5 | 3.3 | 0.47 | 4.3 | 4.43 | 237 |
| 145 | 4.2 | 430 | 12.5 | 4.6 | 0.53 | 4.4 | 3.68 | 224 |
| 175 | 3.9 | 291 | 11.8 | 6.0 | 0.61 | 4.8 | 3.80 | 153 |
| 190 | 4.4 | 254 | 10.0 | 4.1 | 0.70 | 5.5 | 3.38 | 180 |

By application of an external magnetic field normal to the sample surface, the temperature dependence of the second critical magnetic field $B_{C2}(T)$ was measured. From the linear part of the temperature dependence of $B_{C2}(T)$ near $T_C$ we estimated the electronic diffusion coefficient $D = -\frac{4k_B}{\pi e}\left(\frac{dB_{C2}}{dT}\right)^{-1}$. From, $B_{C2}(0)$, assuming a Werthammer-Helfand-Hohenberg dependence[27] in the dirty limit, the correspondent value of the coherence length $\xi(0) = \sqrt{\frac{\Phi_0}{2\pi B_{C2}(0)}}$ is estimated. For increasing sputter current $I_{Sp}$ (increasing Nb content), both $D$ and $\xi(0)$ increase (see Table I) but are close to typical values obtained for NbN films on sapphire[28].

**C. Superconducting energy gap**

To determine the superconducting energy gap value, we performed THz spectroscopy of our NbN films in the frequency range 0.1 – 1.2 THz at temperatures down to 4.2 K. Using a set of tunable, monochromatic THz sources and a Mach-Zehnder interferometer, we measured both the transmission and phase shift of the THz radiation that passes through the thin film samples[29]. To avoid the birefringence of the substrate, we consecutively irradiate the sample with polarization along the two main axes of the R-plane $Al_2O_3$ substrate[30]. Transmission and phase spectra exhibit pronounced Fabry-Perot resonances caused by the substrate and modified by the NbN thin film. From reference measurements on a bare substrate, we determine the optical properties of the R-plane $Al_2O_3$ and use them to describe the spectra of the thin film samples: we simultaneously fit their transmission and phase data, assuming that the properties of the film can be described by a BCS superconductor following the formalism of Mattis and Bardeen[31]. This fit is performed with the superconducting gap $\Delta$ and the critical temperature $T_C$ as only fit parameters whereas the normal-state scattering rate and the plasma frequency, which also enter the fit, were determined from Drude fits to spectra taken above $T_C$. The temperature dependence of Δ fits well to the numerical calculation according to ref. 32. From the energy gap value at zero temperature, we can calculate the superconducting coupling parameter $\Delta/k_B T_C$, which does not change significantly with $I_{SP}$. The average value is 1.91, which is close to what has been measured by tunnel spectroscopy on 3.5-10 nm thick NbN films on sapphire[33] and indicates a strongly coupled superconductor. For further calculations, we use the gap values directly measured at 4.2 K, which is the same temperature as the operation temperature of SNSPD detectors (see Table I).

## III. DETECTOR PATTERNING AND EXPERIMENTAL SETUP

On each film, nine nanowire detector structures were patterned with a combined process of electron-beam- and photolithography as well as reactive ion etching in $SF_6$ plasma. By inspection with SEM, structures with defects or constrictions were sorted out and the line-width $w$ of the wire was confirmed to be within 5 nm of the nominal value of 100 nm with a gap between the lines of 60 nm. The detector area filled by the nanowire was approximately 4x4.2 µm². The detector elements are embedded in coplanar 50 Ω impedance matched layout of contact pads.

The $T_C$ values of the patterned samples show the same relative behavior on $I_{Sp}$ as the films but are about 1.3 K lower due to the proximity effect[34] that becomes apparent in nanostructures. The critical current $I_C^m$ was measured as the current at which the voltage across the sample surpasses a threshold defined by the measurement accuracy. At 4.2 K, the average critical currents of the samples are between 13.8 and 23.4 µA. The measured critical current is compared to the theoretical depairing critical current value $I_C^d$, calculated according to Bardeen:

$$I_C^d = \frac{4\sqrt{\pi}(e^\gamma)^2}{21\zeta(3)\sqrt{3}} \frac{w(\Delta(0))^2}{eR_S\sqrt{D\hbar k_B T_C}} \left[1 - \left(\frac{T}{T_C}\right)^2\right]^{\frac{3}{2}} \quad (2)$$

where $\zeta(3) = 1.202$ and $\gamma = 0.577$. The $I_C^d$-values were further corrected for the fact that the numerically computed current in the dirty limit differs from the analytical Bardeen current[35]. The critical current of the samples follows this temperature-dependence close to $T_C$ well, but for temperatures below ~0.85$T_C$ it stays below the depairing value (see inset in Fig. 3) similar to what has been reported in ref. 4. The deviation of $I_C(T)$ from (2) at low temperatures is due to penetration and movement of vortices resulting in energy dissipation in our superconducting nanowire with a width larger than 4.4 $\xi$[36]. At 4.2 K, the experimental critical current $I_C^m$ is already well below the calculated depairing current $I_C^d$, limiting the maximum ratio of bias current to $I_C^d$ that can be realized in the experiment. The ratio $I_C^m/I_C^d$ evaluated for all samples at 4.2 K is shown in Fig. 3 and continuously increases with the increase of $I_{Sp}$.

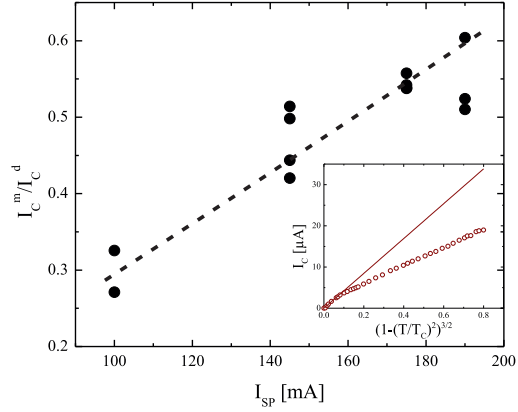

FIG. 3. (Color online) Ratio of the measured critical current of each sample to the calculated depairing critical current at 4.2 K in relation to the sputter current $I_{SP}$ at which the NbN film, of which the detector was structured, was deposited (dashed line is to guide the eye). The inset shows the temperature-dependence of the critical current of one of the samples in direct comparison with the Bardeen critical current (solid line).

Photon count rates of SNSPD response in a broad spectral range were measured at 4.2 K. The detector was mounted with conductive silver paste on a solid brass sample stage that was coupled to the liquid He bath by copper contacts. The light from a broad-band halogen lamp was passed through a monochromator to select the desired wavelength $\lambda$ between 400 and 1100 nm and then coupled into a multimode fiber. The fiber end was 4 mm above the sample surface, so that the light spot was much larger than the detector area and the intensity distribution was homogeneous across the sample position. The position of the fiber end can be readjusted at low temperatures to compensate cool-down drift. The effective radiation power $P_{opt}$ on the detector area was evaluated by a separate measurement in the detector plane of the total optical power with a sensitive photo diode and the intensity profile obtained with a CCD camera. $P_{opt}$ varied with wavelength; for 650 nm the maximal power used for the measurements resulted to approximately 5 pW on the detector area, for other wavelengths the power was below this value. There was no measurable influence of radiation power on $I_C^m$ and the count rate scaled linearly with $P_{opt}$ over more than two decades around the value used for the measurements. The samples were biased at $I_B$ by a battery-powered low-noise current source. The high frequency output signal was amplified at room temperature and then sent to a pulse counter.

The dark count rate (*DCR*) was measured by blocking the fiber vacuum feed-through. The *DCR* value increases exponentially with the bias current (Fig. 4). In the subsequent measurements, the *DCR* of the respective $I_B$ was then subtracted from the evaluated count rates. The count rate under radiation was measured in the whole spectral range at different bias currents $I_B$. The detection efficiency (*DE*) is calculated as the number of counts divided by the number of photons incident on the detector area. The accuracy of the detection efficiency determination in these measurements is ~4 percentage points.

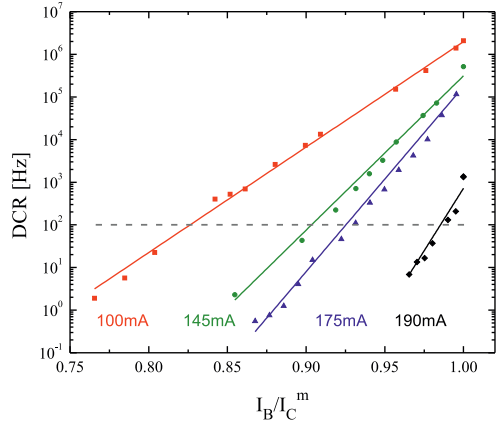

FIG. 4. (Color online) Typical dark count rate (DCR) behavior on relative bias current for one selected detector per each film. The sputter current of the respective film is given below each line. Solid lines are exponential fits of the DCR with respect to $I_B/I_C^m$. The grey horizontal bar marks the threshold of 100 s$^{-1}$.

## IV. RESULTS AND DISCUSSION

### A. Dark count rates

The values of the dark count rates close to $I_C^m$ vary slightly from detector to detector made from the same NbN film due to almost unavoidable differences in the detector mounting onto the sample stage that lead to different cooling conditions[37]. Fig. 4 shows typical *DCR* dependencies on reduced bias current $I_B/I_C^m$ for one detector per NbN film deposited at different $I_{Sp}$. The maximum *DCR* values are reduced almost by three orders of magnitude from the film deposited at 100 mA to the one deposited at 190 mA. Moreover, the *DCR* starts to have significant influence only at higher $I_B$ for films with reduced nitrogen content: A *DCR* of 100 s$^{-1}$ is reached at 0.83 $I_B/I_C^m$ for the 100 mA-deposited film but only at 0.98 $I_B/I_C^m$ for the 190 mA-deposited film. Dark count events are caused by vortices penetrating into the superconductor and dissipating power when moving across the nanowire. They are introduced by thermal excitation over the potential barrier at the edge of the wire[38]. Thus, *DCR* scales exponentially with Boltzman's factor $\exp(-U_{pot}/k_BT)$, where $U_{pot}(I_B)$ is the potential barrier for vortex penetration. Exponential fits to the data (solid lines in Fig. 4) show a continuous increase of the slope for films with reduced N content, indicating an enhanced potential barrier for vortex penetration in those films.

### B. Spectra of the detection efficiency

The spectrum of the detection efficiency shows for all samples a nearly flat, wavelength-independent behavior for small λ. This corresponds to the regime of hot spot detection[3], where the energy of the photon is large enough to create a normal-conducting belt that spans the cross-section of the wire. Above a certain cut-off wavelength $\lambda_C$, which corresponds to the minimum photon energy required for a hot spot formation, the detection efficiency starts to drop. In this regime, the photon can only be detected by photon-assisted vortex nucleation. The achieved maximum detection efficiencies on the hot spot plateau were between 10-16%. The differences in DE between the films are smaller than the accuracy of the measurement, thus we cannot confirm a significant improvement of detection efficiency by variation of the chemical composition in this wavelength range.

However, the cut-off wavelength (compared at the same $I_B/I_C^m$ ratio) appears to be at much longer wavelengths for the films with higher Nb content. Thus, for a given wavelength in the infrared range ($\lambda>\lambda_C$), those detectors have a strongly improved detection efficiency. To quantify this shift, a clear definition of $\lambda_C$ is required. However, in the spectra of the detection efficiency, the hot spot plateau is not completely flat. It even increases slightly due to the influence of the wavelength-dependent absorptance of the NbN film[4,28], which might itself be dependent on the chemical composition of the film. To eliminate these unknown influences and evaluate $\lambda_C$, we need to determine the intrinsic detection efficiency (*IDE*) of the NbN film. The detection efficiency *DE* obtained by the measurements is the product of *IDE* with the absorptance (*ABS*) of the meander structure and geometrical influences (*GEO*) like the filling factor of the detector area

$$DE = GEO \times ABS \times IDE$$

In contrast to the other two factors, the value of *IDE* depends on the applied bias current: The higher $I_B$, the farther $\lambda_C$ is shifted into the infrared region[4]. By normalizing the $DE_1$ of a low bias current, for example $I_{B,1}=0.85 I_C^m$, to the $DE_2$ acquired at high bias current $I_{B,2}=0.95 I_C^m$, we effectively retain a normalized intrinsic detection efficiency

$$\frac{DE_1}{DE_2} = \frac{IDE_1}{IDE_2} \frac{ABS}{ABS} \frac{GEO}{GEO},$$

where the influence of GEO and ABS factors out. Normalized spectra for detectors made from films with different compositions are shown in Figure 5. The shift to higher wavelengths for films with reduced nitrogen content can now be clearly observed.

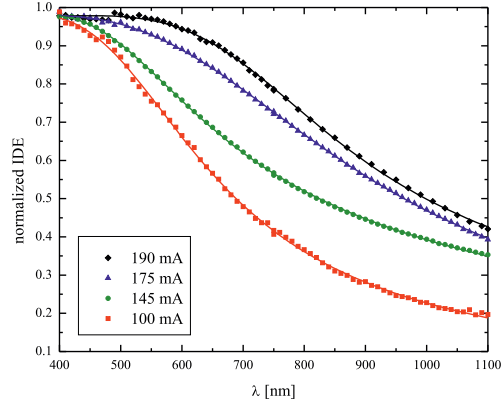

FIG. 5. (Color online) Normalized spectra of detection efficiency of $I_B=0.85\ I_C^m$ to $I_B=0.95\ I_C^m$ for four selected detectors made from different NbN chemical composition as indicated in the legend. The solid lines show the fitting function that was used to determine the cut-off wavelength $\lambda_C$ according to equation 4.

The complete experimental spectral dependence[15] of IDE can be formally best described by

$$IDE = \left(1 + \left(\frac{\lambda}{\lambda_0}\right)^p\right)^{-1} \quad (3)$$

where $\lambda_0$ is interpreted as the cut-off wavelength and the exponent $p$ describes the power-law decrease of the efficiency in the near infrared range. We have to note that the cut-off wavelength defined in such a way need not to coincide with the $\lambda_C$-values in the hot-spot model. The reason is that $\lambda_0$ is affected by non-perfectness of the meander structure and the particular transition between the hot-spot and vortex assisted detection mechanism while $\lambda_C$ only sets the red boundary for the hot-spot detection mechanism in a geometrically ideal nanowire. By fitting the normalized spectra (solid lines in Fig. 5) with the expression

$$\frac{IDE_1}{IDE_2} = \frac{1+\left(\frac{\lambda}{\lambda_{02}}\right)^{p2}}{1+\left(\frac{\lambda}{\lambda_{01}}\right)^{p1}}, \quad (4)$$

the parameters $\lambda_0$ and $p$ for two bias currents can be extracted. The $\lambda_0$-values were determined in this way for various $I_B$ and for several samples from each film. To compare $\lambda_0$-values of detectors from different films to model predictions, $I_B$ is normalized to the individual $I_C^d$ value of each detector. Figure 6 shows the resulting $\lambda_0$ values, where data obtained on samples made from the same film are displayed with the same symbol shape. The spread of $\lambda_0$ for samples from the same film can be caused by individual features in the meander structures like constrictions or turns of the ~85 μm long nanowire. All detectors seem to show the same behavior: as the ratio $I_B/I_C^d$ increases, $\lambda_C$ increases. However, the films with higher Nb content can be operated at higher $I_B/I_C^d$ ratios, because their critical current is closer to the depairing limit, and thus achieve the overall higher $\lambda_C$ values.

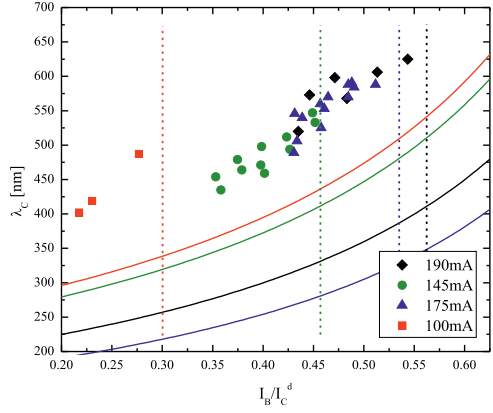

FIG. 6. (Color online) Measured cut-off wavelengths of all samples over the respective ratio of bias current to depairing current at which the measurement was taken. The colors/ symbol shapes mark detectors that were patterned from the same film. Solid lines show the dependence predicted by the hot spot model $\lambda_C = \lambda_{C,0}(1-I_B/I_C^d)^{-1}$, where the $\lambda_{C,0}$ values have been calculated from the material parameters of each film (see table I). The dashed lines mark the mean values of the $I_C^m/I_C^d$ ratios for each film (see figure 3).

To compare the results with the hot spot model, we use film parameters to calculate for each film the pre-factor $\lambda_C(0)$, which is defined as

$$\lambda_c = \lambda_C(0)\left(1 - \frac{I_B}{I_C^d}\right)^{-1}$$

$$\lambda_C(0)^{-1} = \frac{3\sqrt{\pi}}{4hc\varsigma} \frac{\Delta^2 w}{e^2 R_S} \sqrt{\frac{\tau}{D}}. \qquad (5)$$

This prefactor depends only on the material parameters. For $\varsigma$ and $\tau$ we use values that were determined for NbN thin films of comparable thickness $\varsigma = 0.43$ and $\tau = 7$ ps[39]. The theoretical dependencies of $\lambda_C$ on reduced current $I_B/I_C^d$ are shown by the solid lines in Figure 6. Although variations of $\lambda_C$ and $\lambda_0$ with the relative current are pretty similar for all films, there is a systematic offset in experimental $\lambda_0$-values that is most likely due to the difference in defining $\lambda_C$ and $\lambda_0$. When compared at the same relative current, experimental cut-off values are almost the same for different samples from each film and only slightly differ from film to film. The largest cut-off values were found for the film grown at 190 mA. Contrary, the hot-spot model predicts a well pronounced minimum in $\lambda_C(0)$ for the samples from the film grown at 175 mA and the highest $\lambda_C(0)$-value for the film grown at 100 mA. Consequently, the theoretical dependence of the 100 mA film (red line) is above all other lines and would principally yield the highest cut-off values if it were possible to bias at a high $I_B/I_C^d$ ratios. However, the samples from this film had in fact the smallest $I_C^m/I_C^d$ ratio, which marks the limit of the applicable bias current (dashed red line). Thus, despite having the highest $\lambda_C(0)$-value, it cannot achieve competing cut-off values.

The $I_c^m/I_c^d$ ratio shows a clear dependence on the material composition (Fig. 3). For the films with lower content of Nb it is less than 0.3. The ratio grows continuously with increasing sputter current and reaches ≈0.6 for Nb-rich samples. This behavior can be explained by the dependence of $I_c^m$ on the geometry of the SNSPD meander-structure[8,9,10]. An increase in the coherence length ξ (with increase of Nb content, see Table I) results in the decrease of the w/ξ ratio and, consequently, in a weaker current crowding in the sharp 180° turn of the meander line that causes weaker suppression of $I_c^m$ with respect to the depairing current. The suppression is less pronounced for the film deposited at 190 mA. It has the largest ξ(0) and the smallest $T_c$ that results in the largest ξ(4.2K) among all studied NbN films.

One more reason for the discrepancy between experimental cut-off values and model predictions could be our assumption that all films have the same temperature independent efficiencies and time scales (ζ,τ) of the quasi-particle multiplication process. It is not impossible that the mean free path of quasi-particles and the energy gap, which are different for different content of Nb, affect these parameters. Another important simplification is that we used temperature independent electron diffusion coefficients to describe the quasi-particle diffusion below the transition temperature.

## V. CONCLUSION

We performed a systematic and detailed investigation of a series of superconducting nanowire single-photon detectors made from 4 nm thick NbN films deposited at different conditions. By varying the sputter current in vicinity of transition from nitrogen to niobium phase of reactive magnetron discharge, NbN films with continuous variation of their superconducting and normal state properties have been fabricated. The increase of the residual resistance ratio, the electron diffusion coefficient, and the decrease of the resistivity values of films deposited at high sputter current $I_{Sp}$ indicate growth of more uniform NbN films with a reduced amount of defects. This results in an increase of the ratio between the measured critical current and the theoretically estimated depairing current up to ~0.6 for the 190 mA film. Consequently, the SNSPD devices made from NbN films with reduced nitrogen content demonstrate significantly (by almost three orders of magnitude) reduced dark count rate and the cut-off wavelength shifted towards longer wavelengths by more than 25%.

The $I_c^m/I_c^d$ ratio can be identified as the strongest factor influencing on the spectral detection efficiency, as it defines the maximum value of the current-dependent factor in (1) that can be achieved in the detector operation. For the maximum achievable cut-off wavelength of the detector, this limitation is much stronger than the material influence on the $λ_{C,0}$ values predicted by the model. Thus, we can assert that the films with higher barrier for vortex penetration (larger $I_c^m/I_c^d$) are preferable for SNSPD development.

Based on our experimental results we can propose the direction of material search for the development of infrared SNSPD. The optimal material should be uniform, defect free and with high barrier for vortex penetration in patterned sub-micrometer wide structures made from thin films of this material. It is also very important to have mechanically and chemically stable films. The task is very challenging especially for technology. However, in case of optimal detector geometry with minimized suppression of $I_c^m$ with respect to the depairing critical current[8,9,10], superconducting nanowire single-photon detectors with high efficiency in the infrared spectral range could be developed.

## ACKNOWLEDGMENTS

The authors would like to thank E. Hollmann for helpful discussions. The work is supported in part by the DFG Center for Functional Nanostructures under sub-project A4.3.